\documentclass[12pt]{iopart}

\usepackage{graphicx}
\usepackage{bm}% bold math

\newcommand{\bd}{\begin{document}}
\newcommand{\ed}{\end{document}}
\newcommand{\beq}{\begin{equation}}
\newcommand{\eeq}{\end{equation}}
\newcommand{\nid}{\noindent}
\newcommand{\su}{\section}
\newcommand{\ssu}{\subsection}
\newcommand{\sssu}{\subsubsection}
\newcommand{\baR}{\begin{array}}
\newcommand{\eaR}{\end{array}}
\newcommand{\ben}{\begin{enumerate}}
\newcommand{\een}{\end{enumerate}}

\newcommand\bbbc{{\sf I\!\!C}}

\newcommand\cB{{\cal B}}
\newcommand\cH{{\cal H}}
\newcommand\cM{{\cal M}}
\newcommand\cL{{\cal L}}

\bd
\title[]{Does the complex susceptibility of the H\'enon map have a pole
in the upper-half plane ? A numerical investigation.}
%\titlerunning{}

\author{B. Cessac}
\address{INRIA, 2004 Route des Lucioles, 06902 Sophia-Antipolis, France.\\
Institut Non Linéaire de Nice, 1361 route des Lucioles, 06560 Valbonne,
France.\\
Universit\'e de Nice, Parc Valrose, 06000 Nice, France.}

\begin{abstract}
It has been rigorously shown in \cite{CMP} that the complex susceptibility of
 chaotic maps of the interval can have a pole in the upper-half complex plane.
We develop a numerical procedure allowing to exhibit this pole from
time series. 
We then apply the same analysis to the H\'enon map and 
conjecture that the complex susceptibility has also a pole
in the upper-half complex plane.  
\end{abstract}

\pacs{02.70.-c,05.10.-a,05.45.-a}
%\noindent{ \it Keywords\/ Linear response, chaotic systems.\\}
\submitto{\NL}
\maketitle

Consider a dynamical system  $f : \cM \to \cM$, on a manifold $·\cM$
endowed with the Lebesgue measure. Assume that there is a unique $f$-ergodic
probability measure $\rho$ so that $\lim_{T \to \infty}\frac{1}{T} \sum_{t=1}^{T} \phi(f^n(x)) = \int \phi d\rho$
for continuous functions $\phi$ and for Lebesgue-almost every $x \in \cM$. Then,
$\rho$ is called the SRB measure or physical measure (see \cite{Young} for a discussion
on these two terminologies). Assume that when $f$ is perturbed in its neighborhood, the
average value $\rho(A)$ of a given smooth observable $A$ varies differentiably with $f$.
Then the derivative operator is called the \textit{linear response function}.
In the case of uniformly hyperbolic systems, and provided that $f$ and its perturbations
are sufficiently smooth, then the linear response is given by \cite{Ru99}:

\beq\label{drt}
\delta \rho(A) = \sum_{n=0}^\infty \rho(X.\nabla(A\circ f^n)),
\eeq

\nid where $X$ is a (time-independent) perturbation. Ruelle has indeed shown that this series is
convergent for suitable perturbations. In the case of a time-dependent
perturbation $X_t$ this formula extends formally to:

\beq\label{drtt}
\delta \rho_t(A) = \sum_{n=0}^\infty (\kappa_{n} X_{t-n})A.
\eeq  

\nid The operator $\kappa_n : \cH \to \cB^\ast$, where $\cH$ is a suitable space of vector fields and $\cB^\ast$
the dual of a suitable space of functions, is defined by:

\beq \label{kappa}
(\kappa_n X_{t-n})A = \rho(X_{t-n}.\nabla(A\circ f^n)); \qquad n  \geq 0,
\eeq

\nid while, for $n<0$, ($\kappa_n X_{t-n})A=0$ (causality).
It is called the ``response function''. In this case, one recovers the
standard definition of the linear response  in non equilibrium statistical
mechanics ((\ref{drtt}) is a convolution product), and classical results such as fluctuation-dissipation relations
or Onsager theory can be recovered \cite{Ru99}. Moreover, this result does not require the assumption
of closeness to equilibrium.

However, one can find examples of non uniformly hyperbolic systems where
the series (\ref{drt}) does not converge, even   for maps of the interval with absolutely
continuous invariant measures \cite{CMP}. In this case, one considers
the \textit{complex susceptibility} function, $\chi(\omega)=(\hat{\kappa}_\omega X)A$
where $\hat{\kappa}_\omega$ is the Fourier transform
of the response function. It is also convenient to use the variable $\lambda=e^{i \omega}$.
Then, $\chi(\omega)=\Psi(\lambda)$.
Under some conditions on $X$ and $f$
(interval maps), Ruelle proved that $\Psi(\lambda)$ extends to a meromorphic
extension with no pole at $\lambda=1$. He also showed that 
this function has a pole inside the unit disk $|\lambda| <1 $ (i.e.  in the upper-half complex plane for the frequencies
$\omega$) and other poles, corresponding to
the eigenvalues of the Perron-Frobenius operator (Ruelle-Pollicott resonances \cite{Pollicott,RuelleP}), outside the unit disk
(in the lower-half complex plane).
 The existence of a
pole of the complex susceptibility in the upper-half complex plane 
 must not be misinterpreted.
This is not a ``violation of causality'', which has no meaning
in a dynamical system, causal by construction. This simply expresses
that an arbitrary small perturbation of $f$ does not give
a response proportional to the perturbation. According
to Ruelle: ``this might be expressed by saying that $\rho$ is not
linearly stable''. 

This result is  quite interesting and intriguing
for a physicist and two natural questions arise. Does this property exist in larger dimensional
chaotic dynamical systems ? And is it possible to have an experimental/numerical
characterization of the effects of this pole ? The first question is difficult to address from
a mathematical point of view and is still unresolved even for classical
examples such as the H\'enon map. The second one is the main concern of the present paper.
We indeed propose a numerical experimentation protocol allowing to
numerically approximate the complex susceptibility (\ref{susc}) and to investigate the numerically
observable effects of having a pole in the upper-half complex plane.

The paper is  organized as follows. After a brief summary of Ruelle's results
for the logistic map (section \ref{RuelleTh}), we describe our numerical methods
in section \ref{NumMet} and apply it to the logistic map where one can use the guidelines of 
Ruelle's mathematical results as
a validation (section \ref{Log}). Then, we apply it to the H\'enon map for the standard parameter
values and conjecture that this map could also exhibit a pole in the
upper-half complex plane (section \ref{Henon}).

\su{Reminder of Ruelle's results.}\label{RuelleTh}

The dynamical system defined by the logistic map:

\beq\label{f}
f(x)=1-2x^2, \quad x \in [-1,1],
\eeq

\nid has an absolutely continuous ergodic measure with density :

\beq\label{rhoinv}
\rho(x)=\frac{1}{\pi\sqrt{1-x^2}}.
\eeq

The linear response (\ref{drt}) reads, in this case:

\beq\label{drt1D}
\delta \rho(A) = \sum_{n=0}^\infty \int X(x) \frac{d}{dx}A(f^nx) \rho(dx),
\eeq  

\nid while the complex susceptibility reads:

\beq\label{susc}
\chi(\omega)=\Psi(\lambda)=
\sum_{n=0}^\infty \lambda^n \int  A'(f^n x) (f^n)'(x)X(x)\rho(dx)
\eeq 

\nid with $\lambda=e^{i\omega}$.
Ruelle has shown the following results \cite{CMP}.
Consider the variable
change $x=\sin(\frac{\pi}{2} y)=\varpi(y)$
which maps $f$ on $g(y)=1-2|y|$ [tent map]. Then the density $\rho$
is mapped onto the density $\sigma_0(y)=\frac{1}{2}$. 
In this variable, we have :

\beq\label{PsiL0}
\Psi(\lambda)=\sum_{n=0}^\infty \lambda^n \int_{-1}^1 (\cL^n_0 Y)(s)B'(s)ds
\eeq

\nid
where $B=A \circ \varpi$, $Y(y)=\sigma_0(y) \frac{X(\varpi(y))}{\varpi'(y)}$
and $\cL_0\Phi(y)=\Phi\left(\frac{y-1}{2}\right)-\Phi\left(\frac{1-y}{2}\right)$.

Consider first a square integrable holomorphic perturbation $X \equiv X_0$ that \textit{vanishes} at $\left\{-1,1\right\}$.
Then,  the series  
(\ref{PsiL0})  has a meromorphic extension in $\bbbc$ and has no pole inside the unit disc.
Moreover, it is holomorphic at $\lambda=1$ (resp. $\omega=0$ ). 
Indeed, one has, using integration by parts:

\beq \label{PsioL}
\Psi(\lambda)=
-\sum_{n=0}^\infty \lambda^n \int_{-1}^1 (\cL^n Y'_0)(s)B(s)ds,
\eeq

\nid where  $\cL$ is the Perron-Frobenius operator:

\beq\label{PF}
\cL\Phi(y)=\frac{1}{2}\Phi\left(\frac{y-1}{2}\right)+\frac{1}{2}\Phi\left(\frac{1-y}{2}\right)
\eeq

In the space of analytic functions, and in the basis of monomials $\left\{y^m\right\}$, $m=0,1,2, \dots$ this operator
is represented by an infinite triangular matrix. The eigenvalues are on the  diagonal.
 They are of the form $1/4^n, n=0,1 \dots $. 
The corresponding right eigenfunctions are (Bernoulli) polynomials
such that $\sigma_0(y)=\frac{1}{2}$ (invariant density corresponding to the eigenvalue $1$);

\beq \label{sigma1}
\sigma_1(y)=-\frac{1}{3}-2y +y^2
\eeq

\nid (eigenvalue $\frac{1}{4}$), etc... This simple argument can be found
in \cite{Gaspard} pp 123-125. This result is also a direct
consequence of \cite{BaladiRugh}. 

 Therefore, for  a perturbation $Y'_0(y)=\sum_{k=1}^\infty C_k \sigma_k(y)$, 
and for an observable $A$ such that the integral $\int_{-1}^1 Y'_0(s)B(s)ds$
is finite, one finds easily that the susceptibility
has poles $\lambda_k^{-1}, k \geq 1$, outside the unit disk.
For example, if $Y'_0(y)=\sigma_1(y)$, the series (\ref{PsioL})
reads: 

\beq\label{psi0}
\Psi(\lambda) =- \sum_{n=1}^\infty \left(\frac{\lambda}{4}\right)^n  \int_{-1}^1 \sigma_1(s)B(s)ds
\eeq

 This case
is used as a benchmark for the numerical procedure
developed in the next section.
The corresponding perturbation $X_0$ reads, in the variable $x$:

\beq\label{X0num}
X_0(x)=\pi\sqrt{1-x^2}\left[1-\frac{2}{3\pi}\arcsin(x) - \frac{4}{\pi^2}\arcsin^2(x)
+ \frac{8}{3\pi^3}\arcsin^3(x)\right]
\eeq

\nid where  $X_0(\pm 1)=0$. 
\\

Ruelle has also shown that there is a
 meromorphic function :

\beq\label{Phimoins}
\Phi_-(z)=\frac{1}{z+1}+w(z),
\eeq

\nid where $w(z)$ is of order $O(z+1)$,
and where $\Phi_-(\pm 1)=0$ such that $\cL_0\Phi_-=\sqrt{f'(-1)}\Phi_-$ ($=2 \Phi_-$ in our case).
The function  $w$ in (\ref{Phimoins}) is therefore a solution of:

\begin{equation}\label{Wz}
w\left(\frac{z-1}{2}\right)-w\left(\frac{1-z}{2}\right)-2w(z)=\frac{2}{3-z}.
\end{equation}

\nid $w$ is not an analytic function. It can be obtained via the recursion:

\beq\label{wrec}
w(z) = \left\{\begin{array}{ccc}
&&w(2z+1)+\frac{1}{2-2z} ; \qquad  z <0;\\
&&-w(1-2z)-\frac{1}{2+2z} ; \qquad  z >0;
\end{array}
\right.
\eeq

\nid reminiscent of the Tagaki function discussed by Gaspard  in the context
of the multi-baker map \cite{Gaspard}. Returning to the variable
$x$, the function $\Phi_-$ corresponds to :

\beq\label{Xm}
X_-(x)=\pi\sqrt{1-x^2}\left[\frac{1}{\frac{2}{\pi}\arcsin(x)+1} + w(\frac{2}{\pi}\arcsin(x)+1)\right].
\eeq

The corresponding  series $\Psi_-(\lambda)=
\sum_{n=0}^\infty \lambda^n \int \rho(x) A'(f^{n} x) (f^{n})'(x)X_-(x)dx$,
converges if $|\lambda|\sqrt{|f'(-1)|} < 1$,
and, in its domain of convergence, has the form $\frac{G_A}{1-\lambda|\sqrt{|f'(-1)|} }$
where $G_A$ is a constant depending on $A$. The complex susceptibility
has therefore a pole at $\lambda=\frac{1}{\sqrt{f'(-1)}}=\frac{1}{2}$, inside the unit disk 
(resp. $\omega=i\log(2)$).
Therefore, the series $\Psi_-$ diverges for real frequencies $\omega$.

\su{The numerical method.}\label{NumMet}

The numerical method used here has been introduced by Reick in \cite{Reick} and independently in 
\cite{JAB1,JAB2,JAB3}. It is purely heuristic and  there is no mathematical proof
of its validity, though a discussion with some analytic developments
can be found in  \cite{Reick} and \cite{JAB3}. Thus, 
in this section, we shall use formal expressions.   

Consider a time-dependent perturbation of $f$ of type\footnote{Note that this way of perturbing avoids
the introduction of $f^{-1}$ in the computation
of the response function as found e.g. in \cite{Ru99}. It allows one to write convolution formulae
and the Fourier transform in the standard form they have for continuous time systems.
This was the main motivation for this choice in \cite{JAB3} and
we pursue along these lines in the present paper.} :

\beq\label{SDP}
x_{t+1}=f\left[x_t+\epsilon X_t(x_t)\right].
\eeq

The variation of the average value of  $A$, at time $t$, is formally
given by:

\beq\label{suscgen}
\Delta\rho_t(A)=\epsilon \sum_{n=-\infty}^t \int  A'(f^{t-n}x)(f^{t-n})'(x)X_n(x) \rho(dx)
+ O(\epsilon^2).
\eeq

\nid which corresponds to (\ref{drtt}) to the linear order and up
to a factor $\epsilon$. Note that the term $O(\epsilon^2)$ is  not uniform in $t$.\\

Consider now perturbations of type $X_t(x) \equiv X(x)e^{-i\omega t}$ where $\omega$
is real. Then:

\beq\label{RepLin}
\Delta\rho_t(A)=\epsilon e^{-i\omega t} \chi(\omega)+ O(\epsilon^2),
\eeq

\nid where  $\chi(\omega)$, the complex susceptibility for $X(x)$, is given by eq. (\ref{susc}).
Then, at the linear order in $\epsilon$:

\beq
\chi(\omega)=\frac{\Delta\rho_t(A)e^{i\omega t}}{\epsilon}.
\eeq

Since $\chi(\omega)$ does not depend on time one may write:

\beq\label{chiT}
\chi(\omega)=\frac{1}{\epsilon T} \sum_{t=1}^T \Delta\rho_t(A)e^{i\omega t}
\eeq

The time-dependent SRB state $\rho_t$ 
involves an average over initial conditions in
 the distant past \cite{Ru99}.  One can argue (see \cite{Reick} Appendix A)
that the above 
average over $t$ makes the average over initial
 conditions unnecessary. The idea is therefore to replace $ \Delta\rho_t(A)$ by $A(x'(t))-A(x(t))$ 
where  $x'(t)$ is a typical trajectory of the perturbed system,
with a perturbation $X(x)e^{-i\omega t}$  
and $x(t)$ is a typical trajectory of the unperturbed system. Then, one
computes:

\beq \label{chinum}
S_T(\omega,x) = \frac{1}{\epsilon T} \sum_{t=1}^T \left[ A(x'(t))-A(x(t)) \right]e^{i \omega t}
\eeq

The hope is that $S_T(\omega,x) \to \chi(\omega)$ as $T \to \infty$, where the limit is independent of $x(0)$.
  There is no mathematical proof of the above statement.
   Note that $T$ must be taken so large that roundoff errors play their role of selecting the limit to be the SRB state. 
On practical grounds, as discussed in \cite{Reick}, this procedure is numerically
reliable provided that
$\epsilon T$ is sufficiently large and $\omega T >> 1$ (basically
one uses $\epsilon T \omega >>1$). These conditions have been checked in the simulations performed in the present paper.
Since $S_T(\omega,x)$
is expected to give a fairly good approximation of $\chi(\omega)$, the Fourier
transform of the linear response, when $T$ is sufficiently large,
one also expects that the inverse Fourier series of $S_T(\omega,x)$, called $R(t)$ (where
we drop the $T$ and $x$ for simplicity)
gives a fairly good approximation of the linear response (\ref{drtt}).\\

\nid\textbf{Remarks}

\ben

\item The quantity obtained by
this procedure is,  \textit{stricto sensu}, not the \textit{linear} response but
the \textit{total} response since one can apply it for arbitrary large $\epsilon$.
As a test, one must a posteriori check that the susceptibility is independent 
of $\epsilon$ on a certain range of small $\epsilon$ values [e.g.
it does not vary if one replaces $\epsilon$ by $2\epsilon$].

\item In the case of the logistic map,
any typical trajectory of (\ref{f}) approaches the 
points $\pm 1$ within a distance of order $\epsilon$
with a characteristic time of order $\frac{1}{\sqrt{\epsilon}}$. 
In a numerical simulation where $\epsilon$ is small
but finite this arises often, especially if one respects the condition
$\epsilon T\omega >> 1$. However,
in this case, a perturbation $X(x)e^{-i\omega t}$ can push
the trajectory out of the interval $[-1,1]$ leading
to an exponential divergence.
To avoid this effect 
we have used  a perturbation  $X(x,t)=X(x)(1 + e^{- i \omega t})$
where $X$ is positive about $x=-1$
and  negative about $x=+1$, so that the perturbation is always
directed inside the interval $[-1,1]$ whenever $x=\pm 1$.  

\een

\su{Application to the logistic map.}\label{Log}

\ssu{A benchmark.}

Let us first consider the case where the complex susceptibility is well defined
on the real axis (for the frequency $\omega$) by the  series (\ref{susc}).
More precisely, we consider  a  perturbation, corresponding to the Ruelle-Pollicott resonance $\frac{1}{4}$.
 In the  variable $x$ this perturbation is given by eq. (\ref{X0num}).
For the observable $A$ one can use any choice such that the integral $\int_{-1}^1 \sigma_1(s)B(s)ds$ in ( \ref{psi0})
is finite. We choose  $A=X_0$. The complex susceptibility reads $\Psi(\lambda)=-\frac{C}{4-\lambda}$
where $C>0$ is a constant.\\

\begin{figure}[ht]
\begin{center}
\includegraphics[height=8cm,width=7cm,clip=false]{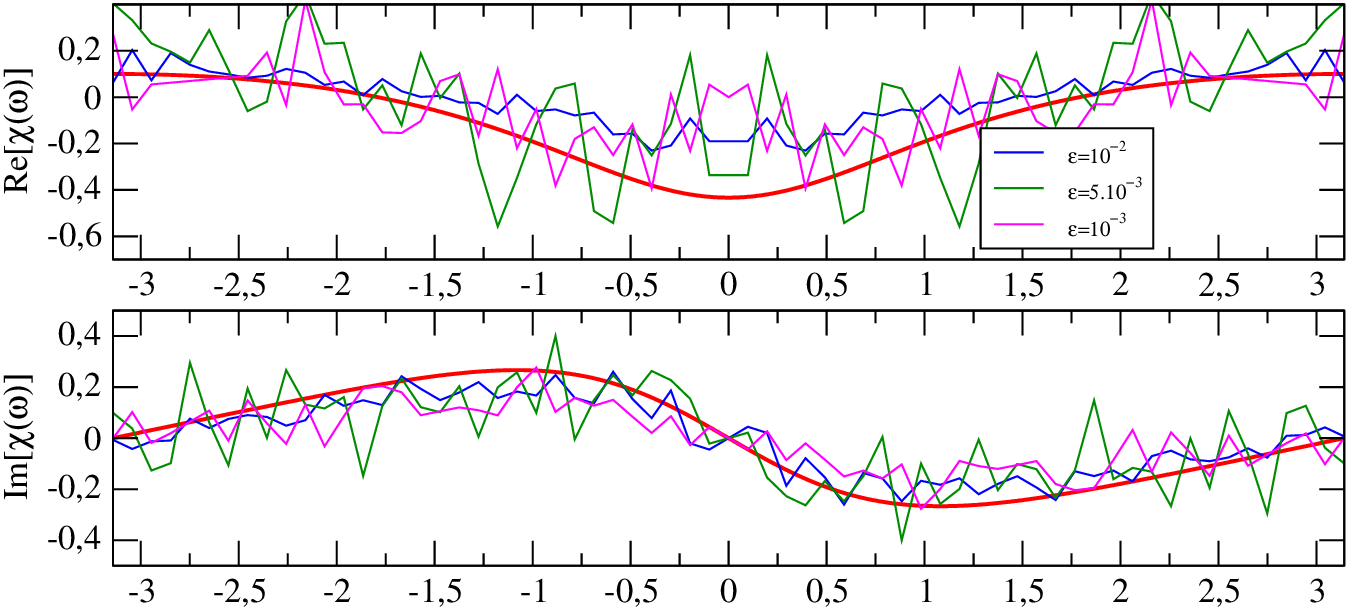}
\hspace{1cm}
\includegraphics[height=8cm,width=7cm,clip=false]{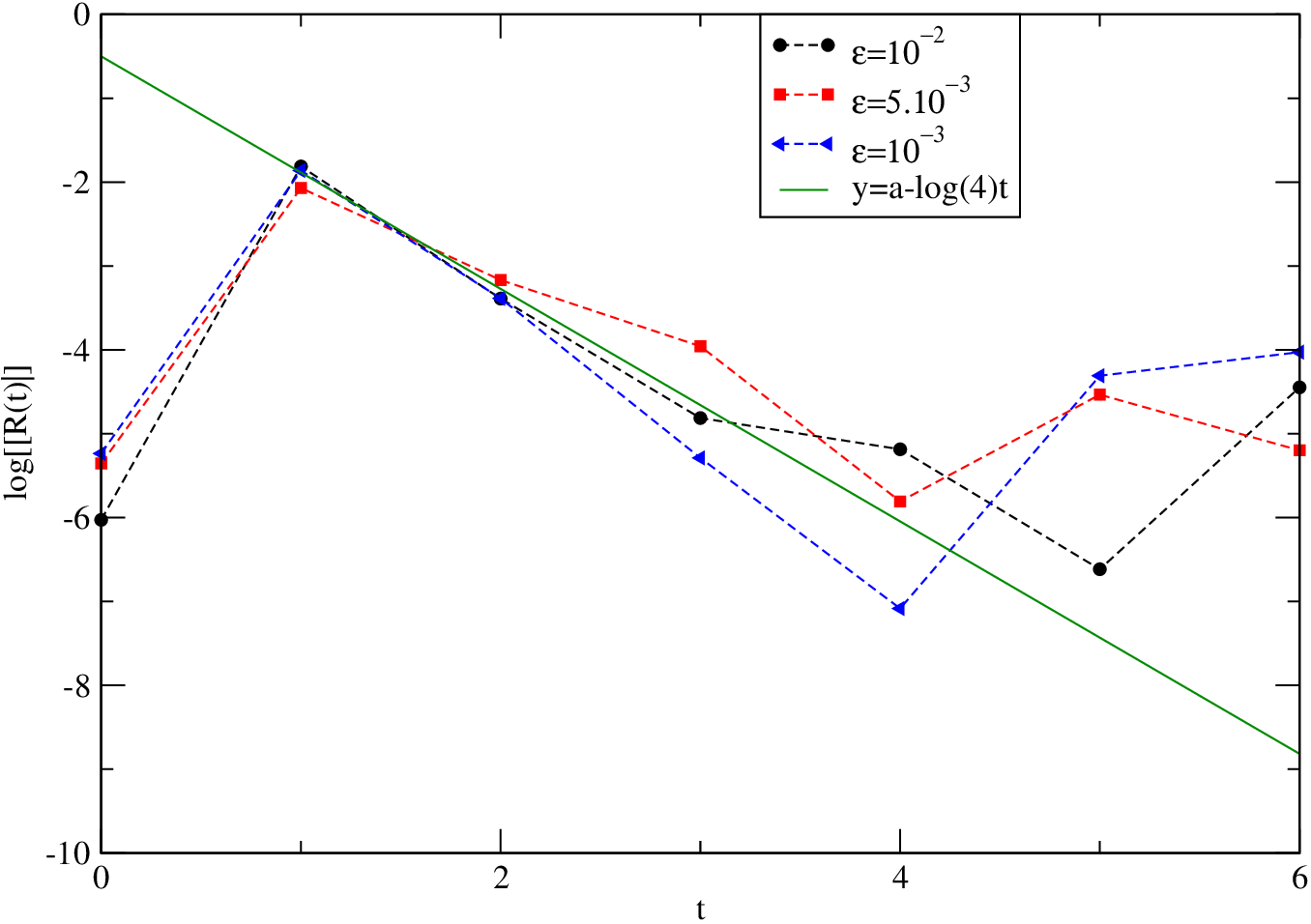}
\vspace{0.5cm}
\caption{(a) Susceptibility for a perturbation (\ref{X0num}) and
an observable also given by (\ref{X0num}), for $\epsilon=10^{-2},5 \times 10^{-3},10^{-3}$. 
(b) Corresponding approximate linear response [denoted by R(t)], in log scale
and theoretically expected curve $y=a-log(4)t$. 
\label{Fres1}}
\end{center}
\end{figure}

In the figure \ref{Fres1}a the complex susceptibility, computed
with this algorithm and compared to the theoretical value, is drawn. 
To generate the real and imaginary parts of the susceptibility
we run two independent simulations. Starting from the same
mother trajectory, we apply in the first case a perturbation $\epsilon \cos(\omega t)$
and compute the corresponding susceptibility, which gives $Re(\chi(\omega))$.
In the second case we apply  a perturbation $-\epsilon \sin(\omega t)$, which gives $Im(\chi(\omega))$. 
Note that the perturbations are therefore real and the trajectories stay on the real axis.
The amplitude of the perturbation
was fixed to $\epsilon=10^{-2},5 \times 10^{-3},10^{-3}$. Averages are performed by computing 
the time average $S_T(\omega,x)$ (\ref{chinum}) for
$N$ initial conditions $x_n, n=1 \dots N$ randomly chosen and then computing
$\frac{1}{N}\sum_{n=1}^N S_T(\omega,x_n)$. This procedure allows us to reduce the
numerical noise and to compute error bars.
We took  $T=10^6,N=10$ for $\epsilon=10^{-2}$;
$T=2.10^6,N=20$  for $\epsilon=5 \times 10^{-3}$ and $T=10^7, N=10$
for $\epsilon=10^{-3}$.
Note that numerical noise is large and
that the resonance curve is
very flat, requiring to have small error bars. This requires
an average over a very long time $T$, roughly given by the condition $\epsilon T \omega >> 1$,
and limits the range of
$\epsilon$ values that one can reach. Note that the condition $\epsilon T \omega >> 1$ is always
violated as $\omega$ approaches zero, whatever the (finite) value of $T$.
This explains the discrepancy observed for small frequencies. 
If one excludes this, the agreement is quite good.
Note also that the experimental curve has a  magnitude
that does not depend of $\epsilon$, 
in this range of $\epsilon$ values, as expected.

   The (approximate) linear response $R(t)$ can be easily obtained by computing the inverse Fourier series
of $\chi(\omega)$ (e.g. by a direct summation). It is drawn  
in Figure \ref{Fres1}b (in log scale for the $y$ axis). Note that
up to numerical noise $R(t)$ is real, as expected. One observes an exponential decay very close
to the theoretically expected decay $4^{-t}$, corresponding to the
mixing rate given by the first Ruelle-Pollicott resonance 
(and also to the characteristic decay rate toward equilibrium, in agreement with the fluctuation-dissipation theorem).
Note that the range of validity for the interpolation is very thin ($\sim [0,4]$). Indeed, rapidly
the perturbation becomes so weak that one measures only the numerical noise.\\

%
%%%%%%%%%%%%%%%%%%%% Susceptibilité res 1
%%%%%%%%%%%%%%%%%%%%%%%%%
%

\ssu{Perturbation $X_-$.}

As we saw, the perturbation (\ref{Xm}) leads to a diverging series (\ref{susc})
for real frequencies. However, when computing the expression (\ref{chinum}) one is not dealing
with a series, but with a finite sum, which diverges 
as $T$ grows. The divergence rate provides useful hints on the presence of the pole.

Consider indeed the formal expression
(\ref{suscgen}) of $\Delta \rho_t(A)$ in the case of the logistic map.
$f^{t-n}$ has $t-n$ zeros and  oscillates rapidly (for large $t-n$) with 
a period $\sim 4^{n-t}$. One can decompose the integral
$\int_{-1}^{+1} \rho(x) A'(f^{t-n} x) (f^{t-n})'(x)X(x)dx$ over intervals
delimited by the zeros of
$f^{t-n}$. The density $\rho$ has strong variations about $\pm 1$  and
small variations in the central part. The contribution
of the first interval (containing $-1$) is given, for large  $t-n$, by :

\beq\label{ExpPole}
\baR{ccc}
&\int_{-1}^{ -1+4^{n-t}} \rho(x) A'(f^{t-n} x) (f^{t-n})'(x)X(x)dx \\
\sim &\frac{4^{t-n}}{\pi} \int_{0}^{4^{n-t}} \frac{A'(f^{t-n}(-1+u))X(-1+u)}{\sqrt{u}}du
\eaR
\eeq

\nid where $x=-1+u$. If $A'(f^{t-n}(-1+u))X(-1+u)$ \textit{does not vanish about $u=0$ ($x=-1$)} and has small variations
on this interval, this contribution is $\sim 4^{t-n} \int_{0}^{4^{n-t}} \frac{1}{\sqrt{u}}du =
2^{t-n}$, which diverges\footnote{If one considers now the contribution of the ``bulk'' [inner intervals] to the integral (\ref{ExpPole})
one can figure out that it does not diverge, essentially because the derivative
$(f^{t-n})'(x)$ has alternating signs and because the density $\rho$
has small variations in the bulk. More precisely, the contribution of the bulk
is provided by a decomposition on the Ruelle-Pollicott eigenfunctions $\sigma_n$ and decays exponentially
due to mixing. Thus, the existence of a pole in the upper-half complex plane is
due to a boundary effect obtained when the perturbation weights the points $\left\{-1,1\right\}$
(where the density diverges). } when $t-n \to \infty$, with a rate exactly given by the pole
$\lambda=\frac{1}{2}$.
Therefore, another good benchmark of our method, in the case where the complex susceptibility
has a pole in the upper-half plane,  consists in computing (\ref{chinum}),
and taking the inverse Fourier series, providing in this way an approximation for $\Delta \rho_t(A)$.
One should see then the exponential divergence with a rate $\log(2)$ (which
is exactly the value of the Lyapunov exponent $\lambda$).

However, there is  another effect that must be taken into account in the numerics.
When  the amplitude of the response, $\sim \epsilon 2^t$, is small
one computes numerically the first term of the Taylor expansion (\ref{suscgen}).
 But when $\epsilon 2^t$ becomes too large the numerical computation
includes also the nonlinear terms and one computes in fact
the complete susceptibility, including nonlinear terms. This one does not diverge
because the trajectories of the perturbed system remain in the interval $[-1,1]$,
 but it  does not
depend linearly on $\epsilon$.\\

Therefore, one expects the following. There is  a time cut-off   
 $t_m(\epsilon) \sim - \frac{\log(\epsilon)}{\log(2)}$
beyond which one does not compute the \textit{linear} response but
the complete response,
including nonlinear effects which saturate the growth of the perturbation.
Then one should observe an exponential growth $2^t$ up to $t_m(\epsilon)$ in the linear response
of the
form  $2^t H_{[0,t_m(\epsilon)]}(t)$,
where $H_I()$ is the characteristic function of the interval~$I$. The corresponding susceptibility
  is:

$$\sum_{n=0}^{t_m}  (2\lambda)^n=
\frac{1-(2\lambda)^{t_m}}{1-2\lambda}$$

\nid which reads, using the frequency $\omega$:

\beq\label{Explth}
\baR{ccc}
&\frac{(1-2\cos(\omega))(1-2^{t_m}\cos(\omega t_m))+2^{t_m+1}\sin(\omega)\sin(\omega t_m)}{5-4cos(\omega)}&\\
+i \ &\frac{2 \sin(\omega)(1-2^{t_m} \cos(\omega t_m))-2^{t_m} \sin(\omega t_m)(1-2\cos(\omega)))}{5-4cos(\omega)}&
\eaR
\eeq

\nid Therefore, it exhibits oscillations due to the cut off $t_m$.
For longer time, mixing should lead to an exponential decay of the response.

\begin{figure}[ht]
\begin{center}
\includegraphics[height=7cm,width=7cm,clip=false]{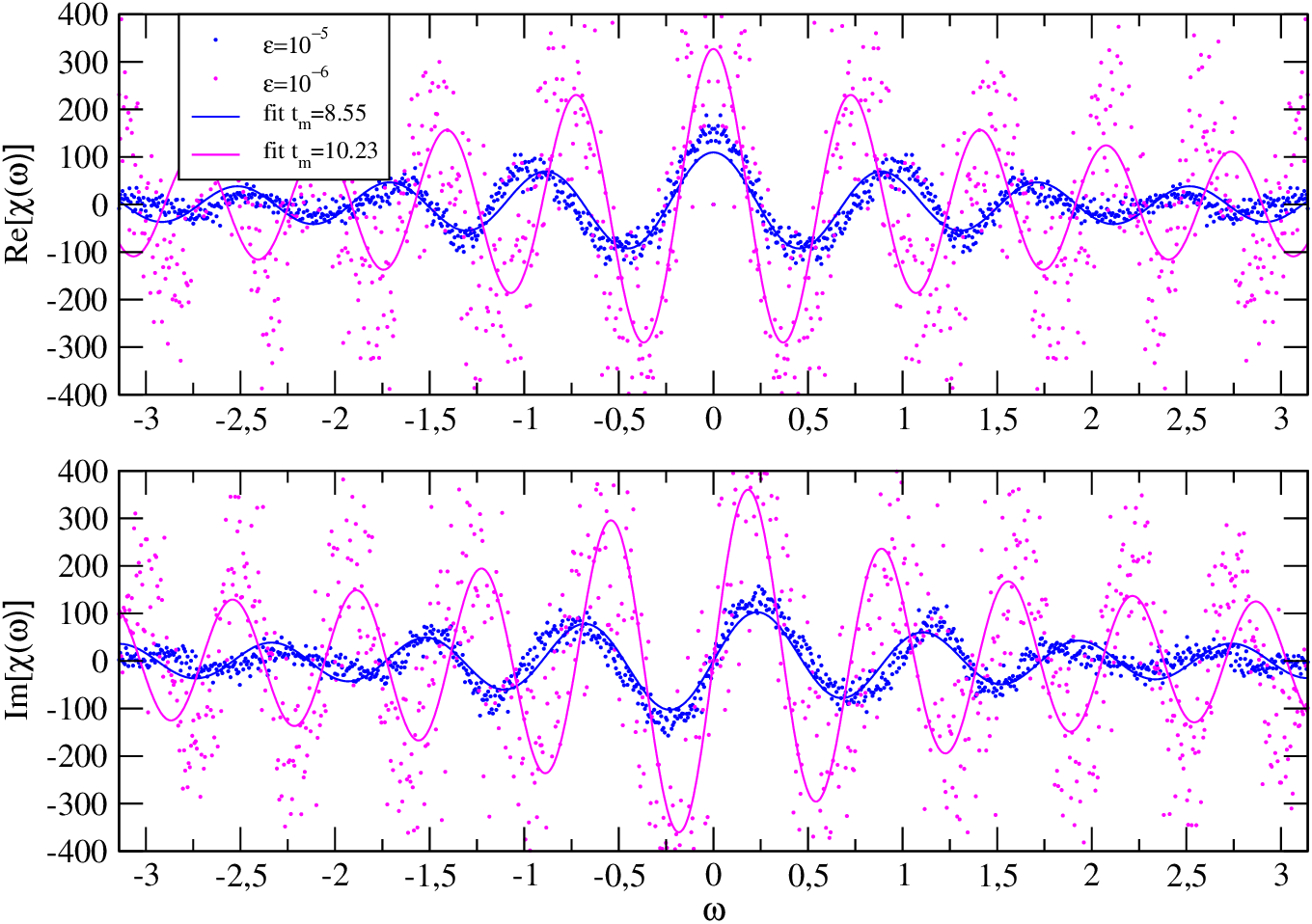}
\hspace{1cm}
\includegraphics[height=7cm,width=7cm,clip=false]{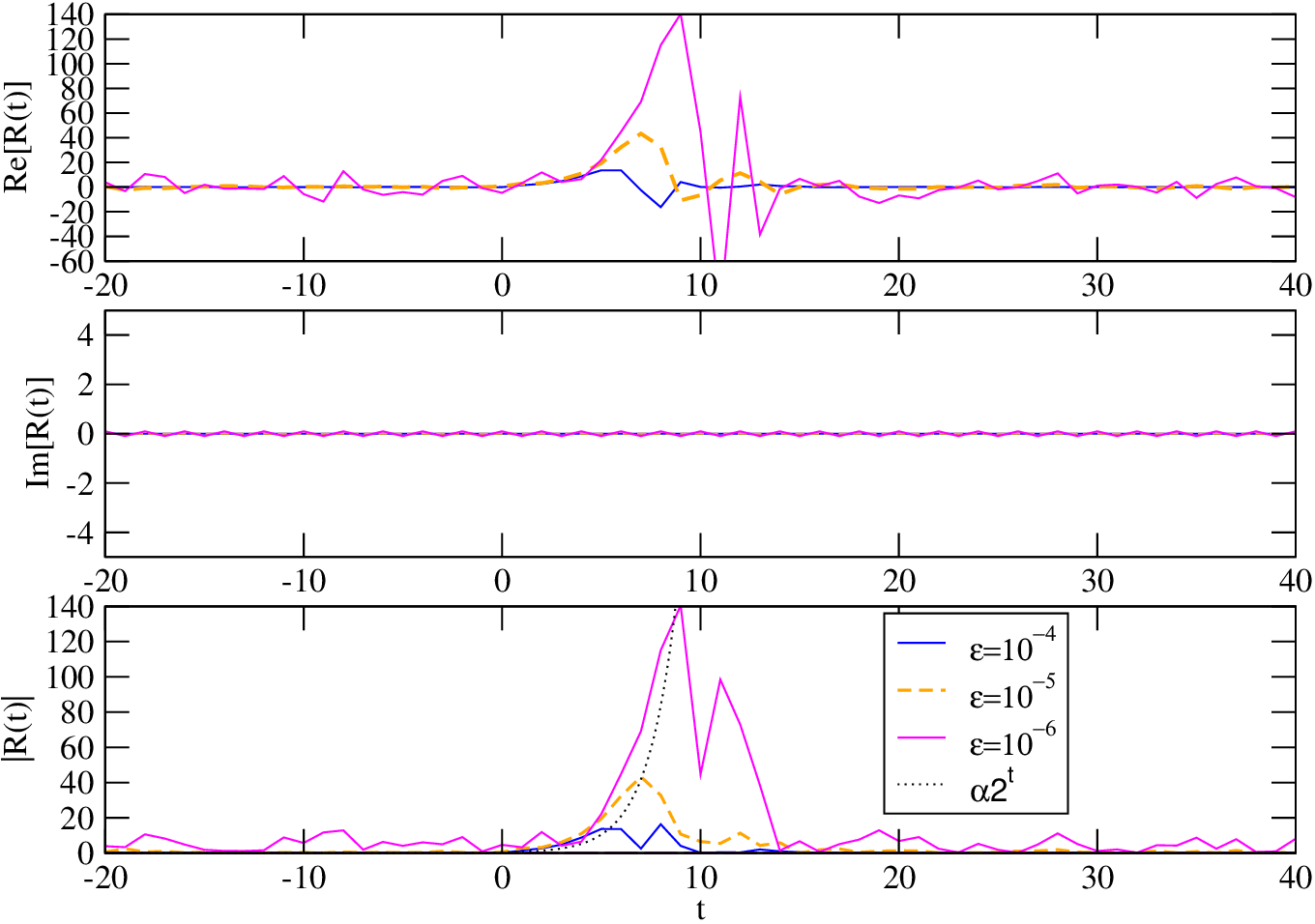}
\vspace{0.5cm}
\caption{(a) Susceptibility for the perturbation 
$X_-(x)$, where $w$ was computed up to order $4$,
and observable $A(x)=x-\frac{x^2}{2}$. The parameter $\epsilon$ takes
the value $10^{-5},10^{-6}$. In full line are drawn the fitting curves obtained
from (\ref{Explth}).
(b)Linear response.
\label{Fres2}}
\end{center}
\end{figure}

To check this we have first computed (\ref{chinum}) for the perturbation (\ref{Xm}) and for the observable
$A(x)=x-\frac{x^2}{2}$ (the derivative of $A$ gives
a maximal value for the term $A'(f^{t-n}(-1+u))$ in equation (\ref{ExpPole})).
The function $X_-$ has been computed with the recursion (\ref{wrec}) up to the
order $4$.  The $\epsilon$ values $10^{-2},10^{-3}$,$10^{-4} (T=10^5,N=400)$,$10^{-5} (T=10^6,N=400)$,$10^{-6} (T=4\times 10^6,N=1600)$ have been considered 
(but only $10^{-4},10^{-5},10^{-6}$ are represented in Fig. \ref{Fres2}a,b for the legibility
of the figure.). We have
used a fit procedure to compare the experimental data with the form (\ref{Explth}).
We have also computed the approximate  linear response. The results are
presented in Fig. \ref{Fres2}a,b.

%
%
%
%
%%%%%%%%%%%%%%%%%%%% Susceptibilité
%%%%%%%%%%%%%%%%%%%%%%%%%
%
%%
%
%
%

\begin{figure}[ht]
\begin{center}
\includegraphics[height=7cm,width=7cm,clip=false]{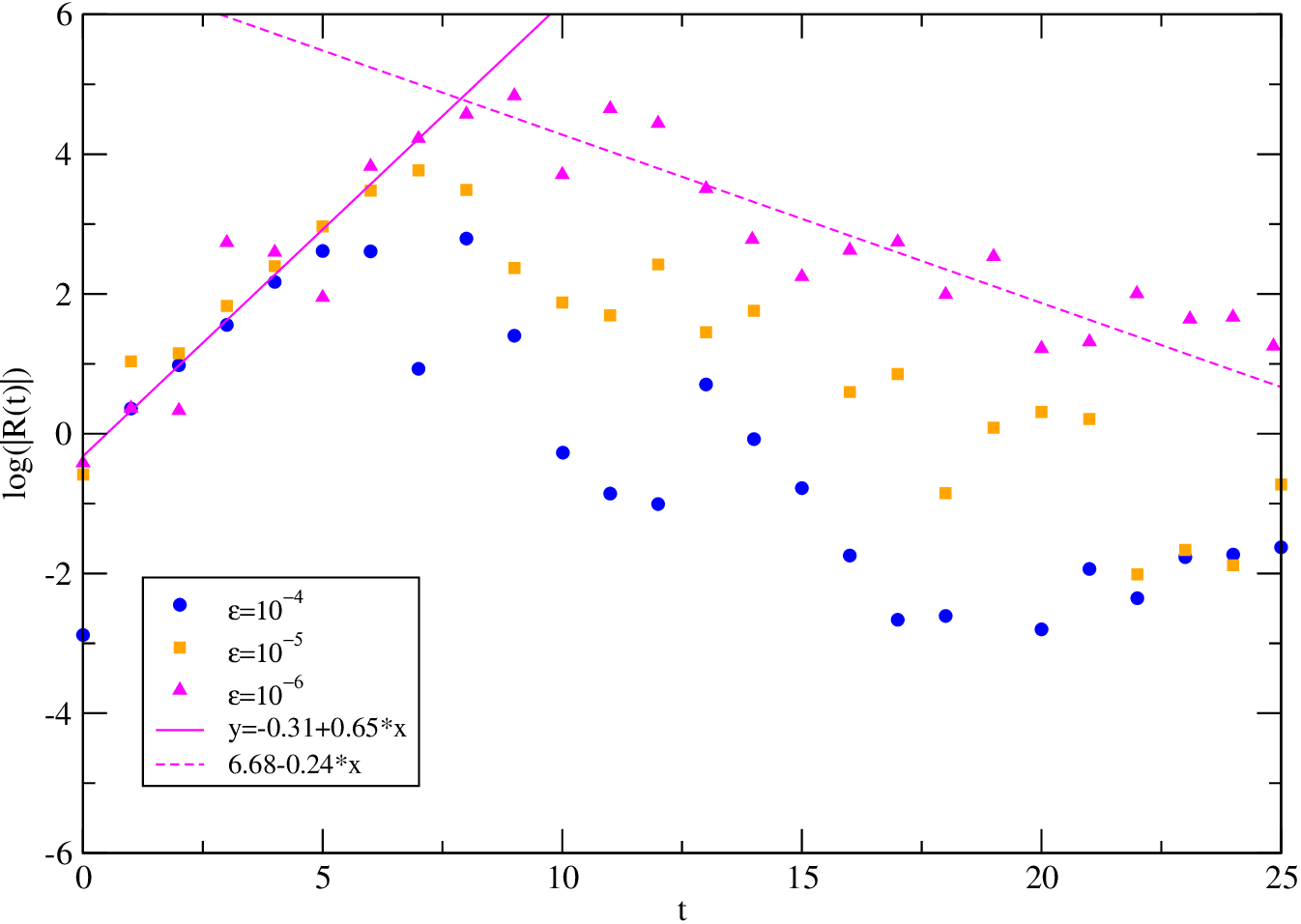}
\vspace{0.5cm}
\caption{Fit of the linear response corresponding to Fig. \ref{Fres2}b.
\label{FInterpolReponseunebossecutoff}}
\end{center}
\end{figure}

One observes indeed a clear dependency with $\epsilon$. Moreover, the real and imaginary part
exhibit the expected oscillations due to the cut-off, with a perfect agreement with the 
form (\ref{Explth}). Taking the inverse Fourier series one obtains the linear response
in Fig. \ref{Fres2}b. One sees clearly the exponential growth $2^t$ and the $\epsilon$
dependent cut-off.  Performing a fit in log scale one obtains (fig. \ref{FInterpolReponseunebossecutoff})
an exponential increase with a rate $0.65$ very close to the expect value $log(2)=0.693$.
After this there is an exponential decay with an approximate rate $-0.24$.  (We have only represented
the fit for $\epsilon=10^{-6}$ but one sees easily in Fig.\ref{FInterpolReponseunebossecutoff}
that the decay rate is similar for $\epsilon=10^{-4},10^{-5}$). This rate
is slower than the first Ruelle Pollicott resonance $-\log(4)=-1.386$. It might be that we are observing a crossover regime
where exponential amplification and exponential mixing are competing. 

%
%
%
%
%%%%%%%%%%%%%%%%%%%% Fit 
%%%%%%%%%%%%%%%%%%%%%%%%%
%
%%
%
%
%

\su{The H\'enon map.}\label{Henon}

The H\'enon map:

\beq
\left\{
\baR{ccc}
x_{t+1}&=& 1-ax_t^2 + y_t\\
y_{t+1}&=& bx_t
\eaR
\right.
\eeq

\nid with $a=1.4,b=0.3$ has a strange attractor with a fractal
structure. Moreover, numerical estimates of the largest Lyapunov exponent,
using e.g. Eckmann-Ruelle algorithm \cite{ER} gives a value $\sim 0.422$.
But the H\'enon map  is not uniformly hyperbolic. There are points with tangencies between
stable and unstable manifolds.  These points may be responsible for phenomena analogous
to those described in the previous section,
for specific perturbations weighting those points. From this observation,
it is conjectured that there may also exist
a pole in the upper-half complex plane (and possibly more complex
singularities \cite{RCom}). However, there is no mathematical result for
this and the form of the perturbations/observable leading to 
such a singularity is not known.       

\begin{figure}[ht]
\begin{center}
\includegraphics[height=9cm,width=7cm,clip=false]{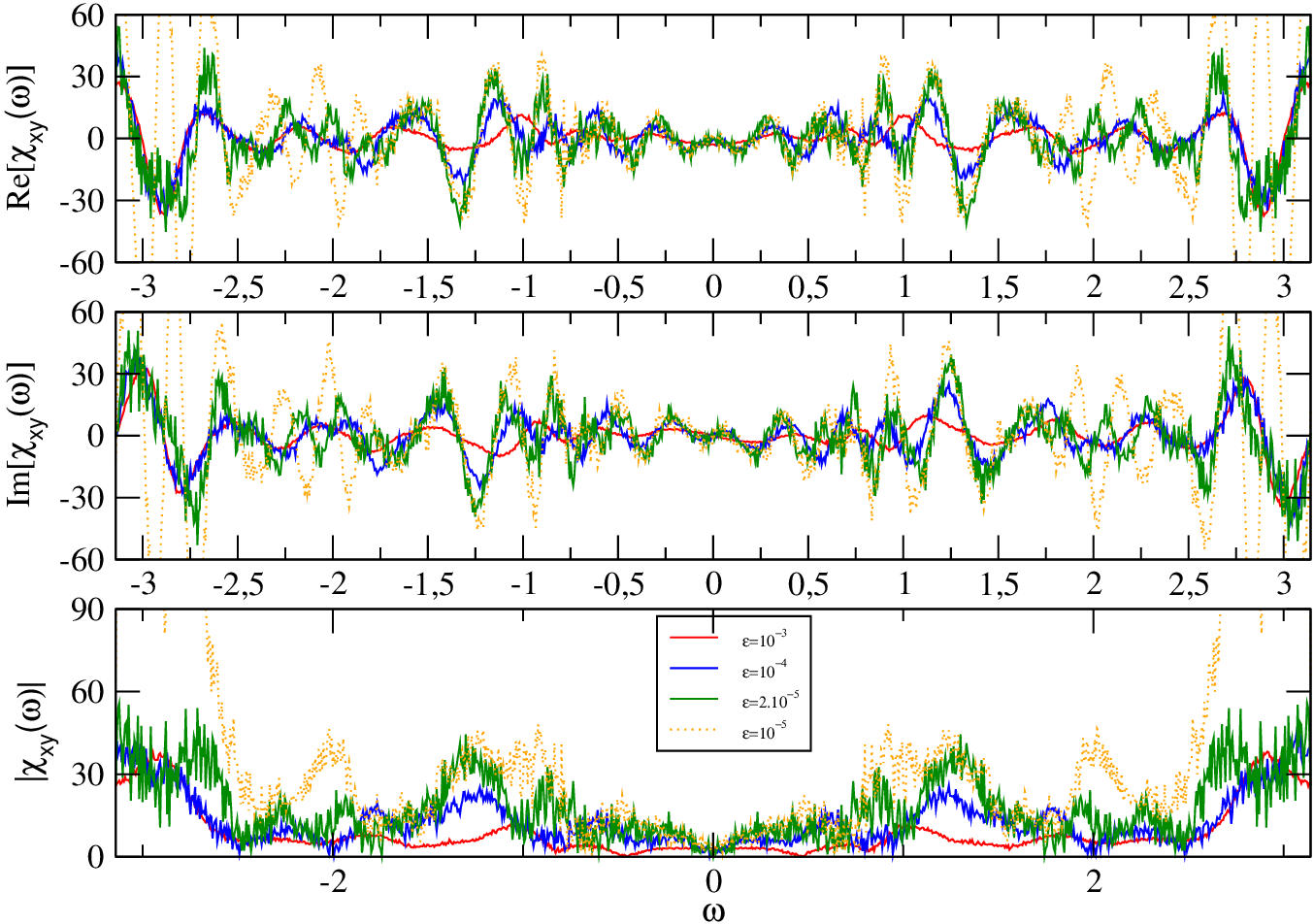}
\hspace{1cm}
\includegraphics[height=9cm,width=7cm,clip=false]{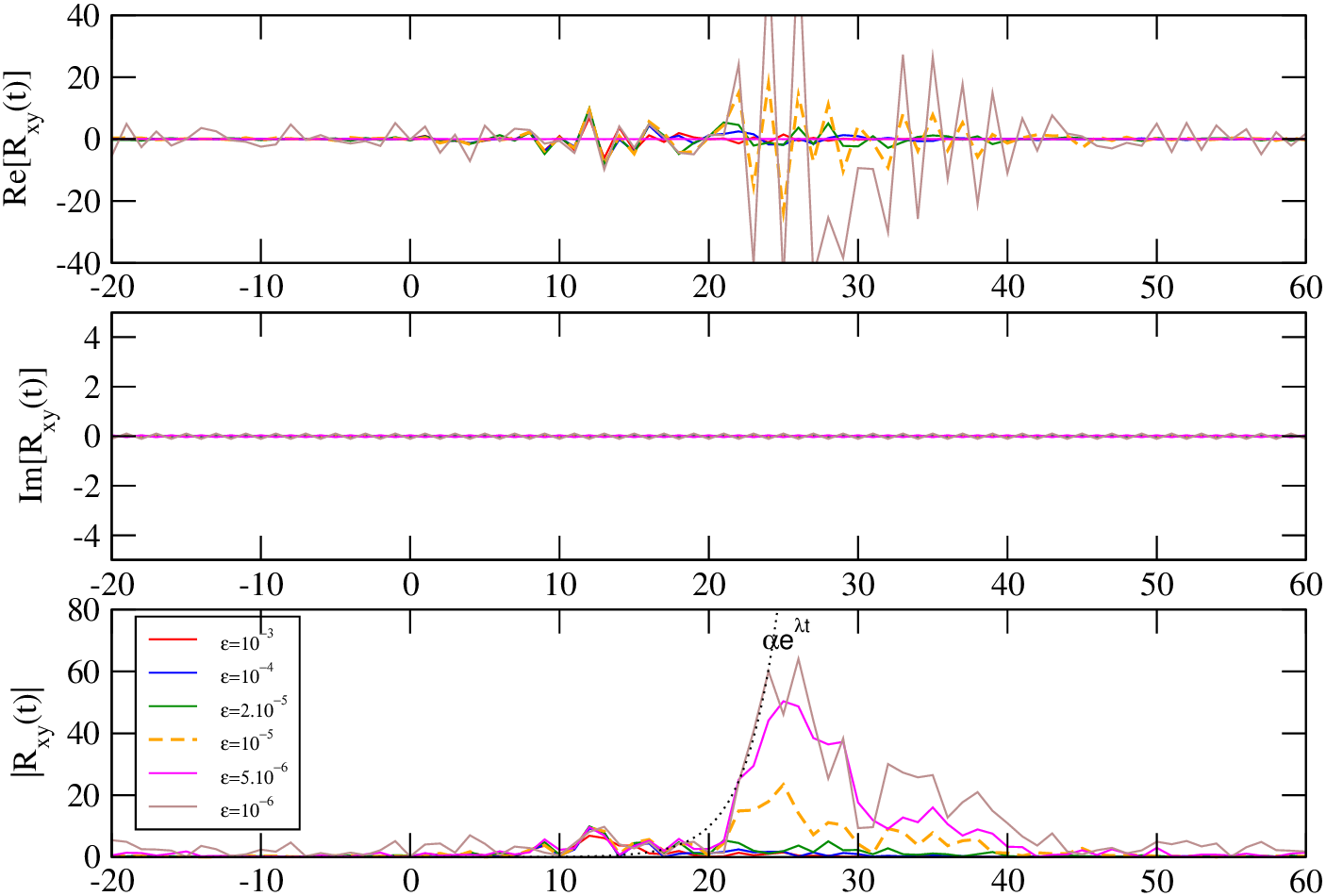}
\vspace{0.5cm}
\caption{(a) Susceptibility for a constant perturbation  $X=\epsilon$
in the direction $y$ for the observable $A(x,y)=x$.
(b) Corresponding linear response.}\label{SusHenon}
\end{center}
\end{figure}

Using the same method as in the previous section, a natural empirical approach
 consists in applying a perturbation $\epsilon e^{i\omega t}$ in one of the directions $(x,y)$
and investigating the effect on the linear response for one of the variables  $(x,y)$. 
As an example,  we have numerically computed the response of the observable $A(x,y)=x$ to a  perturbation
$\epsilon e^{i\omega t}$ in the direction  $y$. Denote by $\chi_{xy}(\omega)$ and $R_{xy}(t)$  the corresponding susceptibility and
approximate response. They are  drawn in Fig. \ref{SusHenon}a,b 
respectively, for $\epsilon=10^{-3} (T=10^5,N=100),10^{-4} (T=10^5,N=100),2 \times 10^{-5} (T=10^6,N=400),10^{-5} (T=10^6,N=1600)$. Note that essentially the same curves are obtained,
up to a phase factor, when perturbing direction $y$ and computing the response of $y$. In particular $|\chi_{xy}(\omega)|=|\chi_{yy}(\omega)|$.

%
%
%%%%%%%%%%%%%%%%%%%% Susceptibilité pour une perturbation constante en x et observable y, modèle de Hénon 
%%%%%%%%%%%%%%%%%%%%%%%%%
%
%
%
%
%
%
%

One observes similar effects as in the previous section. The amplitude 
of the complex susceptibility increases with $\epsilon$ (and this effect is not stabilized, neither
by increasing the time $T$ of average well beyond the criterion $\epsilon\omega T >>1$,
nor by increasing the number of sample trajectories). The linear response has  three parts. 
For short times ($t<20$) there is a small bump for
the modulus of the response: its height does not seem to depend on $\epsilon$. 
The perturbation/observables have a  nonzero projection  onto
the Ruelle-Pollicott modes. It might be that this bump comes from
this part, having a well behaved linear response.
The second part exhibits an exponential increase with a time cut-off depending
on $\epsilon$. This is very similar to the observations made for the
logistic map: exponential amplification of the linear response 
until nonlinearities induce saturation of this effect.
Then, mixing leads to an exponential decay (third part of the curve).

\begin{figure}[ht]
\begin{center}
\includegraphics[height=7cm,width=10cm,clip=false]{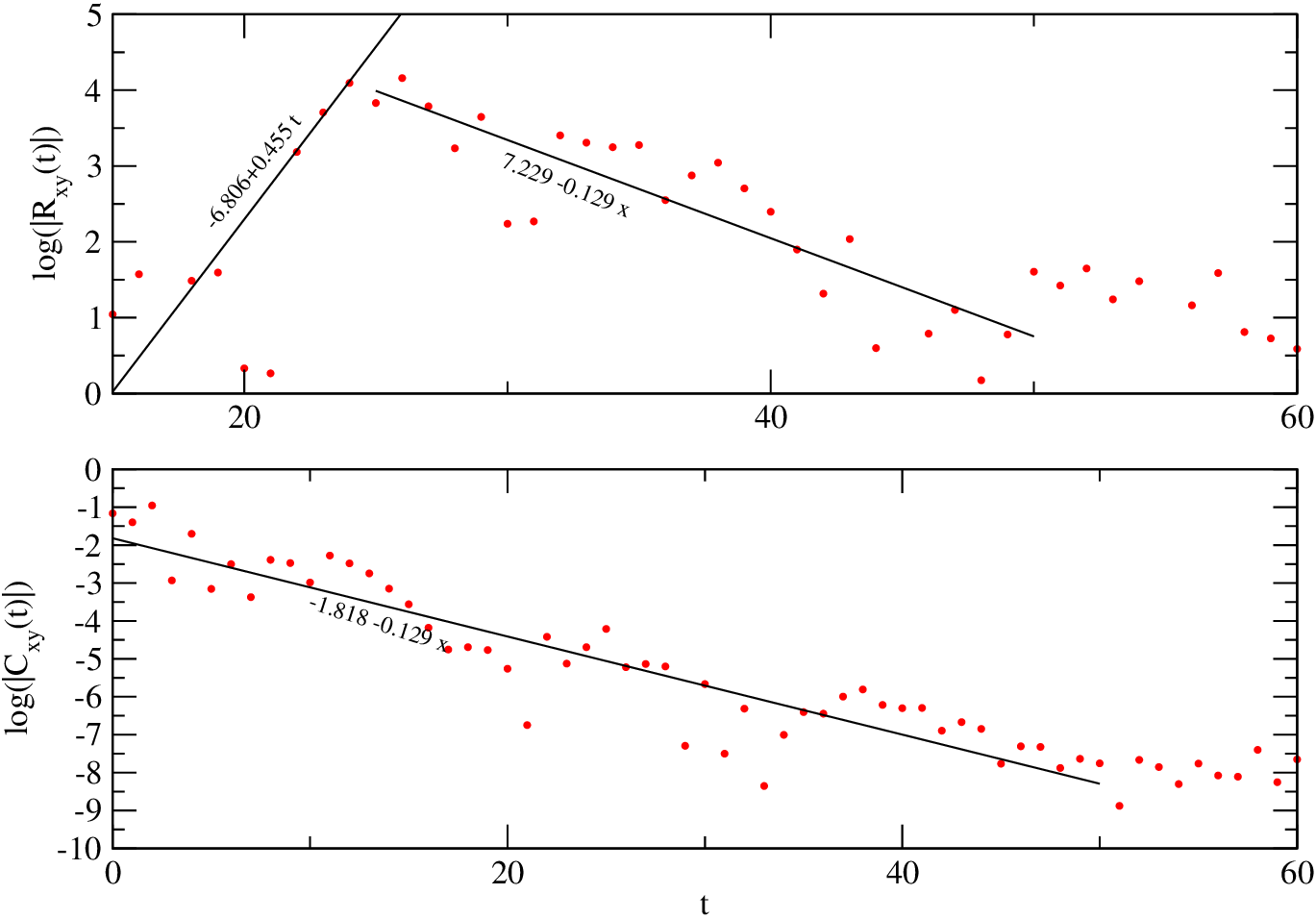}
\vspace{0.5cm}
\caption{(a) Fit of the linear response $|R_{xy}(t)|$, $\epsilon=10^{-6}$ for the H\'enon map.
(b) Fit of the correlation function $C_{xy}(t)$ for the H\'enon map.}\label{Finterpol}
\end{center}
\end{figure}

The exponential instability is well fitted by the curve $e^\alpha e^{\lambda t}$ with $\alpha =-6.80 \pm 0.87$
and $\lambda=0.455 \pm 0.038$ (see Fig. \ref{Finterpol}).
Note that the positive Lyapunov exponent  of the H\'enon map
is $0.42(2)$ for these values of $a,b$.
The decay in the third part is well interpolated by a sum
of  exponential terms.
The dominant term is $\beta e^{\gamma t}$
with $\gamma=-0.129\pm 0.015$.  The value of $\gamma$
is in agreement with the exponential decay of the
correlation function
$C_{xy}(t)=<f^t(x) y>-<x><y>$,
where $<g>=\int g(X) \rho(dX), \ X=(x,y)$ and $\rho$ is the SRB measure ( Fig. \ref{Finterpol}). Thus the last part apparently
obeys the fluctuation-dissipation theorem.
Correlation function has been computed by the standard 
Fast Fourier Transform method (see \cite{NR}, chapter
13.2).

Let us also remark that there are no thin peak in the  susceptibility
(the frequency resolution is $0.00612$) and the thickness of the peaks
does not change with $\epsilon$.
Since the width\footnote{Consider the complex susceptibility in the neighbourhood of a simple
pole $\omega_0=x_0 + i y_0$.  Then $\chi(\omega) \sim\frac{A}{\omega-\omega_0}$. Assume that this pole is close to
the real axis. Set $\omega=x+iy$. On the real axis $|\chi(\omega)|^2 \sim \frac{|A|^2}{(x-x_0)^2+y_0^2}$.
The modulus  is maximal at $x=x_0$, and has value $\frac{|A|}{y_0}$ (resonance peak). 
The width of  the resonance peak is $2|x_1-x_0|$ where $x_1$ is such
that $[\chi(x_1)|=\frac{|A|}{2 y_0}$. Thus, the width is $2y_0$. } of the resonance peak is given by the imaginary part
of the pole, this suggests that the imaginary part of the poles is bounded away from
zero. This leads us to conjecture that there is no pole on the real axis
(for the frequency $\omega$). This conjecture is compatible\footnote{I thank
one of the reviewers for this remark.} with 
the available results (in particular exponential decay of correlations
for many parameters $a$ and $b$ (see \cite{BeYou1,BeYou2,You,WaYou})).

%
%%%%%%%%%%%%%%%%%%%% Interpolation exponentielle, modèle de Hénon 
%%%%%%%%%%%%%%%%%%%%%%%%%
%
%
%
%
%
%
%

\su{Conclusion.}

There is currently an intensive research activity in mathematics dealing with linear response
theory for  interval maps  (see e.g. \cite{BA1,BA2} and reference therein).
In the case of non uniformly hyperbolic maps one can find  examples where the susceptibility has an arbitrary large number of poles
in the upper-half complex plane \cite{JiRu} or where the average of a smooth function
with respect to the SRB measure of the perturbed system $f + \epsilon X$ is not even Lipschitz
at $\epsilon=0$ \cite{BA1}. These results are intriguing for a physicist and raise some natural
questions: can we measure some effects related to these situations?
What is the ``genericity'' of these examples? Do these effects arise in dynamical systems modeling physical
systems ? do they occur in larger dimensions? Answering  these questions at the level of
rigor of mathematics is probably a too formidable task at the current state of research in this field.
One hope is that some hints can be provided by numerical approaches.

In this spirit, we have presented in this work a numerical procedure allowing to
compute the complex susceptibility/linear response. For this, we use an approximation
of the complex susceptibility (\ref{susc}) by a finite sum, which
can be easily computed. In the cases where the series (\ref{susc}) converges
this sum gives a fairly good approximation. When there is a pole in the upper-half
plane a correct treatment of this sum allows us to recover the pole location.
Applying this procedure to the H\'enon model we conjecture that such
a pole could also be present. 

There are several possible developments of the present work.
On mathematical grounds it could be interesting to apply
this procedure to the  cases discussed in e.g. \cite{JiRu,BA1,BA2}
that can be addressed by rigorous methods. (In particular the cases in
\cite{BA1,BA2} should provide examples where the linear response
does not decay exponentially, but does not grow exponentially either.)
An intermediate case between these mathematically tractable cases and
the difficult H\'enon model would be Collet-Eckmann maps. 
Another question is: ``what is there beyond the pole'' ?
In some sense, the pole in the upper-half complex plane may ``hide'' more complex
singularities lying behind it such as cuts. 
 Is it possible to have any (numerical) idea
of which type of singularities are there ? A natural 
way of doing this is to remove the exponential instability by
multiplying the perturbation by a damping factor (or equivalently
to use a complex frequency $\omega$). This is natural from a mathematical
point of view, but tricky in the numerics, because the damping factor
is either smaller than the exponential instability (and the response grows  rapidly inducing nonlinear effects, as we saw)
or it is bigger (and the perturbation becomes rapidly numerically $0$, then  we are measuring short time transients).
One has thus to make small variations of the damping factor around the pole and look
at the changes in the susceptibility/response curve.

On more physical grounds, a natural continuation
of the present work, could be to  consider a lattice of coupled logistic maps,
to apply an harmonic perturbation at some point and look at the induced effects,
in the spirit of this work. This would be one step
towards the investigation the effects of large dimension (``thermodynamic limit'')
on the singularity induced by the microscopic dynamics.

\ack 
I warmly thank David Ruelle for suggesting to me the
present work and for helpful remarks and suggestions.
I thank Jean-Louis Meunier for illuminating advices and Jacques-Alexandre Sepulchre
for helpful discussions.
I think that this paper has been widely improved due to the remarks of the referees
and their positive criticism. I greatly acknowledge them. \\

\ed